# Ultrafast dynamics of surface plasmon nanolasers with quantum coherence and external plasmonic feedback

Dmitri V. Voronine,[1,2*] Weiguang Huo,[1,3] and Marlan Scully[1,2,4]

[1]*Texas A&M University, College Station, TX 77843, USA*

[2]*Baylor University, Waco, TX 76798, USA*

[3]*Xi'an Jiaotong University, Xi'an, Shaanxi 710049, China*

[4]*Princeton University, Princeton, NJ USA*

Spasers have been theoretically predicted and experimentally observed and promise to deliver new exciting nanophotonic and biomedical applications. Here we theoretically investigate ultrafast dynamical properties of spasers with external plasmonic feedback. We consider a spaser both as a nanoscale source and detector of plasmons which could be used to design novel nano-imaging and sensing techniques. We show that, as with conventional lasers, spasers are sensitive to external feedback. However, unlike the lasers, spasers have faster relaxation dynamics which could be used to develop new ultrasensitive near field imaging techniques. We investigate the dependence of spaser relaxation oscillations on feedback parameters and show that quantum coherence can be used to increase the sensitivity to feedback.

# 1. Introduction

Quantum coherent surface plasmon generators (spasers) and nanolasers have recently been proposed[1-4] and experimentally realized as proof-of-principle demonstrations[5-9]. They provide nanoscale coherent sources of light which can improve spatial resolution and sensitivity of laser-based imaging and sensing techniques. Spasers are different from the conventional lasers in their relaxation rates, emission intensities and dynamics. Further experimental and theoretical studies of spasers are needed to improve their performance and to develop the promising applications. To improve the performance we recently proposed the use of quantum coherence in three-level gain media coupled to a silver-nanoparticle spaser and predicted an order of magnitude increase of the steady-state number of plasmons[10]. The quantum coherence can break detailed balance and modify energy level populations of the gain medium. Lasers without inversion (LWI) operate based on similar principles[11-14], and quantum coherence was recently used to increase the power of laser and photocell quantum heat engines[15,16]. However, the effects of quantum coherence on laser dynamics are less understood.

Lasers can have complex dynamics and can exhibit a wide range of dynamical phenomena including relaxation oscillations and chaotic behavior[17]. The dynamics of lasers was previously used to develop techniques for imaging in turbid media using frequency-shifted optical feedback[18] and relaxation oscillations[19]. The latter technique was based on modification of the short-cavity laser relaxation frequency induced by coherent optical feedback from an external target. The laser was used both as a light source and a sensitive detector. Its dynamics was shown to be more sensitive to feedback than its output power for a class B laser. Laser optical feedback techniques were used for coherent microscopy of cells[20,21] and for near field mapping[22,23]. These approaches can be extended to the nanoscale and realized using spasers with an external plasmonic feedback. The external feedback can be provided by plasmons generated by the spaser and scattered off by nearby surface targets, by a second spaser or by additional sources of plasmons placed in a vicinity of the first spaser. High sensitivity of the spaser to external feedback can be used to develop new nanosensors. Ultrafast dynamical properties of surface plasmons may be used to provide feedback on the femtosecond time scale which may be useful for ultrafast nano-optical applications[24]. Therefore, it is important to understand the spaser dynamics and differences from lasers.

There have been only a few studies of spaser dynamics[4,10,25]. Ultrafast spaser dynamics in the stationary and transient pumping regimes were investigated to make the spaser work as an ultrafast quantum nanoamplifier[4]. Spaser relaxation oscillations without the coherent drive were observed in the simulations of spaser dynamics before reaching the cw regime[4]. In another work, the spaser dynamics have been attributed to Rabi oscillations of the gain medium[25]. The dynamics could be controlled by varying the initial excitation conditions. Understanding such complicated and highly nonlinear dynamics will help improve the performance and amplification efficiency of spasers.

In this work we study the dynamics of spasers and their sensitivity to an external plasmonic feedback. We consider the spaser as a nanoscale coherent light source and a photodetector simultaneously and investigate the dependence of spaser dynamics on feedback parameters. We investigate the effects of quantum coherence induced by an external coherent drive on spaser dynamics and show that quantum coherence improves sensitivity. Our results provide insight into the spaser dynamics and can be used towards extending the feedback laser imaging techniques to the nanoscale.

## 2. Results

The schematic of the proposed spaser with external feedback is shown in Fig.1a. A plasmonic metallic nanostructure such as a silver nanosphere is covered by three-level gain medium quantum emitters such as atoms, molecules or quantum dots, which have a ground state |1⟩ and two excited states |2⟩ and |3⟩ (Fig.1b). The nanosphere can be attached to the tip of a scanning probe microscope and used for imaging. States |1⟩ and |3⟩ are coupled by an incoherent pump. The transition |2⟩→|1⟩ occurs spontaneously and is nearly resonant with the plasmon mode of the nanosphere, and is used to transfer energy from the gain medium to plasmons. The plasmonic field of the nanosphere provides an internal feedback to the gain medium. An additional external feedback field can be coupled to the spaser via propagating surface plasmons arriving from external sources or from scattering of the outgoing spaser fields. For example, the nanosphere can be placed in the vicinity of a plasmonic waveguide and can be used to detect external propagating surface plasmons, $\Omega_b{}^{\grave{}}$ (Fig. 1a).

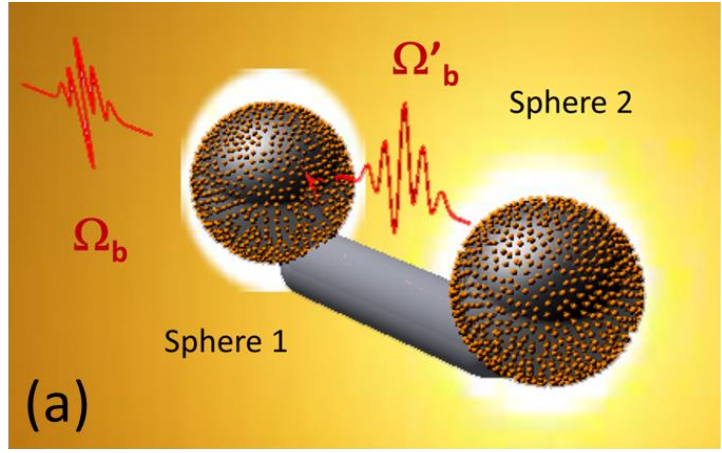

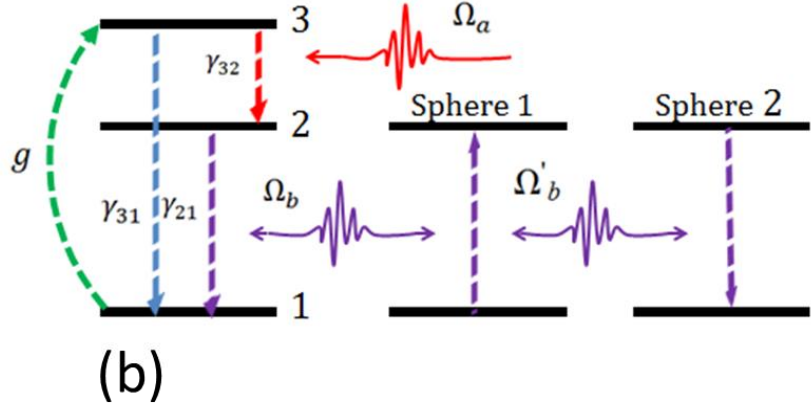

Figure 1. (a) A silver nanosphere spaser (Sphere 1) coupled to an external feedback plasmon source (Sphere 2) with a Rabi frequency Ω'$_b$ via a cylindrical waveguide. (b) Energy level diagram of a spaser consisting of a single plasmon mode (Sphere 1) with external spaser feedback (Sphere 2) coupled to a three-level gain medium driven by an external laser drive, Ω$_a$.

We describe the system using semiclassical theory and density matrix equations. Briefly, we treat the gain medium quantum mechanically and the plasmons and photons classically. We consider the plasmon and photon fields $a_n = a_{0_n} e^{-i\nu_a t}$ and $b_m = b_{0_m} e^{-i\nu_b t}$, respectively, where $a_{0_n}$ and $b_{0_m}$ are slowly varying amplitudes. The Hamiltonian can be written as

$$\mathcal{H}_{int} = \sum_p \{-\hbar\Delta_b^{(p)}|1\rangle\langle 1| + \hbar\Delta_a^{(p)}|3\rangle\langle 3| - (\hbar\Omega_b^{(p)}|2\rangle\langle 1| + \hbar\Omega_a^{(p)}|3\rangle\langle 2| + c.c)\}, \quad (1)$$

where $\Omega_b^{(p)} = -A_n \boldsymbol{d}_{21}^{(p)} \nabla \phi_n(\boldsymbol{r}_p) a_{0_n}/\hbar$ is the Rabi frequency for the spasing transition $|2\rangle \rightarrow |1\rangle$, and $\Omega_a^{(p)} = -\mathbf{E}_m(\mathbf{r}_p)\mathbf{d}_{23}^{(p)}b_{0_m}/\hbar$ is Rabi frequency for the driving transition $|2\rangle \rightarrow |3\rangle$. The summation is over all the *p*th chromophores. $\Delta_a$ and $\Delta_b$ are detunings, defined as $\Delta_a = \omega_{32} - \nu_a$ and $\Delta_b = \omega_{21} - \nu_b$.

The density matrix elements $\rho_{ij}$ satisfy the Liouville-von Neumann equation

$$\dot{\rho}^{(p)} = -\frac{i}{\hbar}[\mathcal{H}_{int}, \rho^{(p)}] - \mathcal{L}\rho^{(p)}, \qquad (2)$$

where $\mathcal{L}$ is the dissipative superoperator. The corresponding time evolution equation for the plasmonic field is obtained using the Heisenberg equation of motion

$$\dot{a}_{0_n} = -\Gamma_n a_{0_n} + i\sum_p \rho_{21}^{(p)} \widetilde{\Omega}_b^{(p)}, \qquad (3)$$

where $\Gamma_n = \gamma_n + i\Delta_n$, $\widetilde{\Omega}_b^{(p)} = \Omega_b^{(p)}/a_{0_n}$ is a single pasmon Rabi frequency. We assume that the Rabi frequencies are the same for all chromophores and omit the index (p) below.

In the numerical simulations we used the following parameters. The detuning of the gain medium from the plasmon mode is $\hbar(\omega_{21} - \omega_n)$=0.002 eV. The external dielectric has the permittivity of $\epsilon_d$=2.25. The nanosphere plasmon damping rate $\gamma_n$=5.3×10$^{12}$ s$^{-1}$ with $\Delta_n$=3×10$^{12}$ s$^{-1}$. The gain medium quantum emitters have the following decay rates: $\gamma_{21}$=4×10$^{12}$ s$^{-1}$, $\gamma_{31}$=4×10$^{10}$ s$^{-1}$, $\gamma_{32}$=4×10$^{11}$ s$^{-1}$; the detunings $\Delta_a$ and $\Delta_b$, and the dephasing rate $\gamma_{ph}$ were set to zero. In all simulations we used the incoherent pump rate, g=8×10$^{12}$ s$^{-1}$. The seed plasmon field at time zero was set to 10$^{-5}$. As an example, we consider a silver nanoparticle of 40 nm radius and the corresponding surface plasmon resonance (SPR) at $\hbar\omega_n$=2.4 eV.

The results of numerical simulations are shown in Figs. 2 – 5. After turning on the pump and before the system can reach equilibrium, it undergoes a sequence of population relaxations leading to a train of short pulses. The physics of these population relaxations is the same as in the conventional lasers[17]. Briefly, when the population inversion reaches a threshold value the system emits a plasmon pulse and the population inversion drops down. When the plasmon field reaches the level below the steady-state equilibrium the population inversion starts increasing again leading to the generation of the second plasmon pulse, and so on. The difference between spasers and lasers are the dynamical time scales which are determined by the relaxation rates which are on the subpicosecond time scale for surface plasmons. The gain medium relaxation rates are also fast due to the coupling to plasmons via the Purcell effect. This leads to faster dynamics of relaxation oscillations. Ultrafast spaser relaxation oscilations without the coherent drive have previously been described by Stockman, M.[4] The spaser dynamics can also be visualized in the phase plane which shows periodic traces in the $n_{21}$ - $N_n$ plot converging to a stable equilibrium point (Fig. 2i). The phase-plane description provides a useful clear view of the relation between the population inversion and number of plasmons.

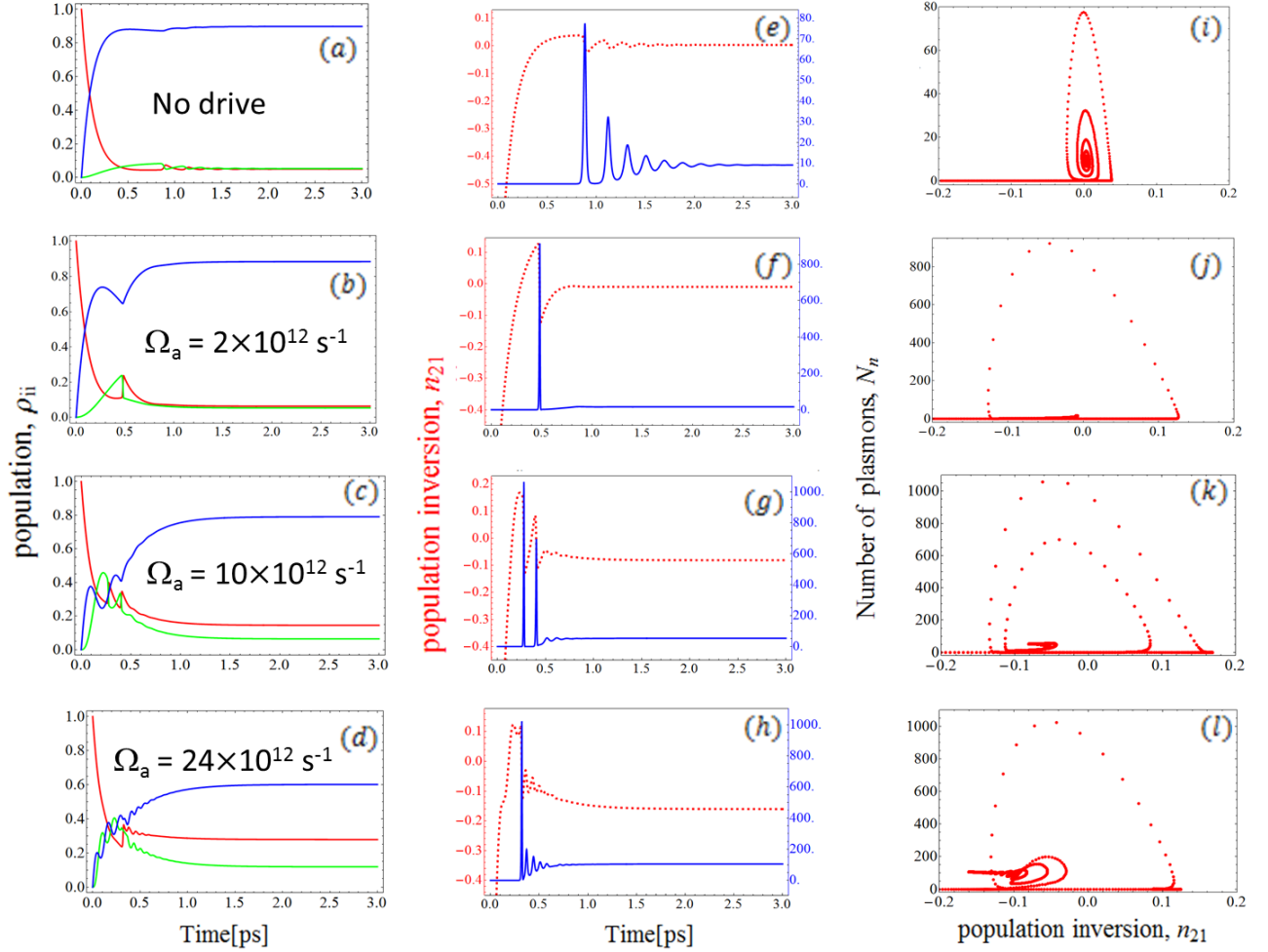

Figure 2. Effects of the drive Rabi frequency, $\Omega_a$, on the spaser dynamics. (a) – (d) Temporal population profiles; (e) – (h) number of plasmons (solid) and population inversion (dashed) plots; (i) – (l) phase plane plots for various values of $\Omega_a = 0$, 2, 10 and $24 \times 10^{12}$ s$^{-1}$.

The plots in Fig. 2 show that as the drive Rabi frequency increases the relaxation oscillation frequency also increases. The first fix spikes appear within approximately 1 ps and 300 fs for $\Omega_a = 0$ and $24 \times 10^{12}$ s$^{-1}$, respectively. For large $\Omega_a$, the first large spike is followed by a train of small oscillations without population inversion. This can be seen in Fig. 2h. As opposed to the case of no drive with a stable equilibrium point in the phase space (Fig. 2i), the stable point of each successive spike shifts to the region of lower population inversion (Fig. 2l). Fig. 2 shows that the spaser dynamics is sensitive to the drive amplitude and can be used as a sensor for imaging the spatial distribution of near fields at the drive resonance frequency. However, the main goal of this work is to investigate the sensitivity of the spaser to the external plasmon feedback with the same frequency as the spaser. Therefore, next we consider the effects of the additional plasmon field on the spaser dynamics. We denote the Rabi frequency of the spaser plasmon field $\Omega_b^0$, and add a new field $\Omega'_b$ due to the feedback from a second plasmon field with a time delay $\tau_f$:

$$\Omega_b(t) = \Omega_b^0(t) + c_f \Omega_b^0(t - \tau_f), \qquad (4)$$

where $\Omega_b^0(t)$ is the Rabi frequency on the spasing transition $|2\rangle \to |1\rangle$ of the first sphere, and $c_f$ is the feedback strength coefficient. In the case of the feedback due to backscattering of the first sphere plasmons, $c_f$ represents the strength of the backscattered plasmon field. In the case of the feedback due to the second plasmon sphere or the second spaser, $c_f$ represents the field strength of the external plasmon source. The feedback may be delayed due to the round-trip plasmon propagation time in the first case, and due to the time-delayed excitation or the one-way plasmon propagation time in the second case. Fig. 3 shows effects of the feedback strength, $c_f$, on the number of plasmons (solid blue) and on the population inversion (dashed red) of the first sphere spaser for the feedback delay time $\tau_f = 1$ ps for a sphere without ($\Omega_a = 0$, left column) and with a drive ($\Omega_a = 24 \times 10^{12}$ s$^{-1}$, right column) for $c_f = 0.1$ ((a), (e)), 0.25 ((b), (f)), 0.5 ((c), (g)), and 0.75 ((d), (h)). The dynamics are modified dramatically by the quantum coherence induced by the drive field and due to the feedback. Faster relaxation oscillations appear as a high amplitude periodic signal in the case with a strong drive. Whereas in the case without a drive, the periodic signal is visible only for a stronger feedback strength of $c_f > 0.5$. The periodic signal may be used to estimate the relaxation oscillation frequency and the feedback delay time and could become a useful experimental tool for studying the plasmon propagation.

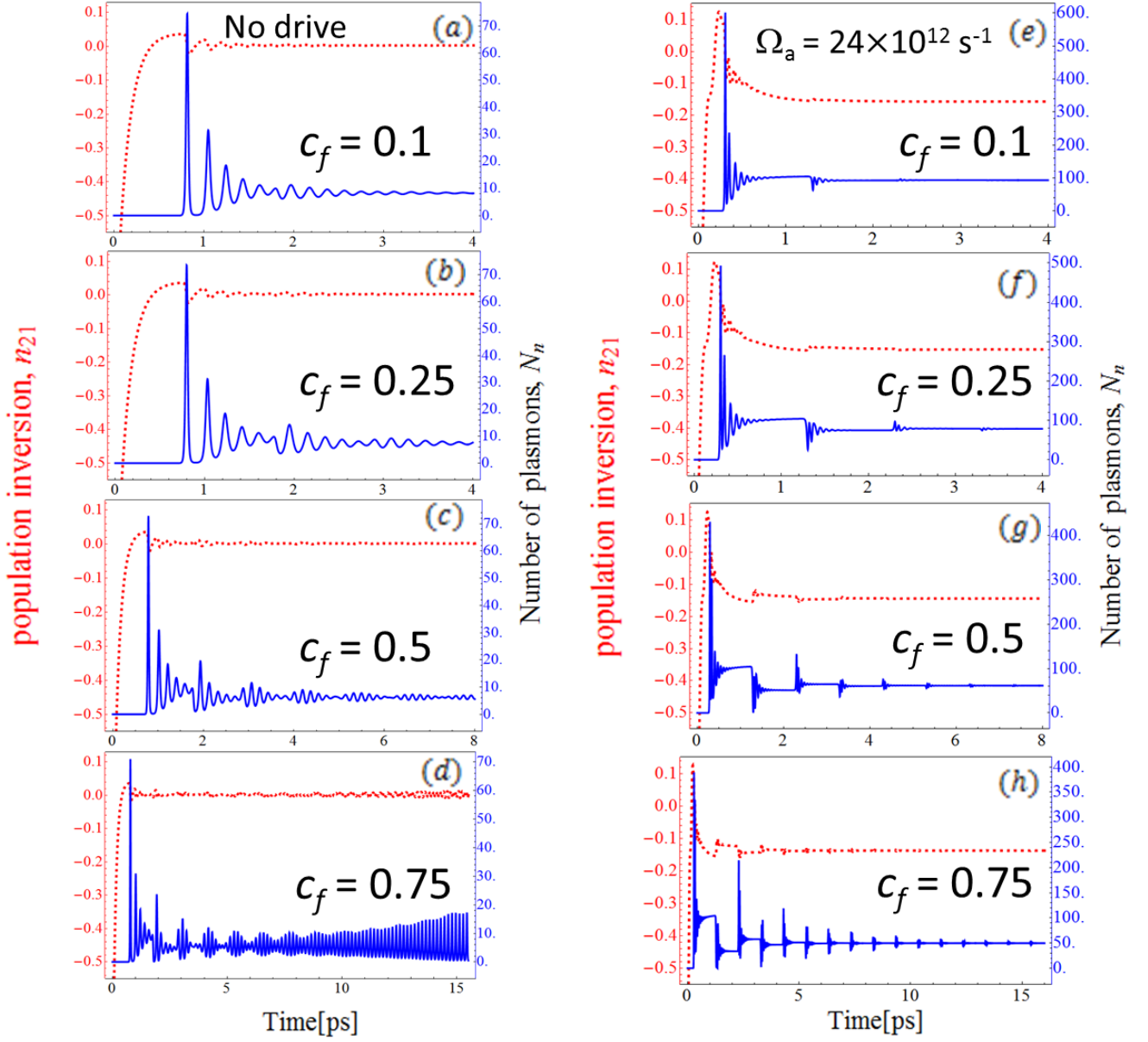

Figure 3. Effects of the feedback strength, $c_f$, on the number of plasmons (solid blue) and on the population inversion (dashed red) of the first sphere spaser for the feedback delay time $\tau_f = 1$ ps for a sphere without ($\Omega_a = 0$, left column) and with a drive ($\Omega_a = 24 \times 10^{12}$ s$^{-1}$, right column) for $c_f = 0.1$ ((a), (e)), 0.25 ((b), (f)), 0.5 ((c), (g)), and 0.75 ((d), (h)).

Next we investigate the effects of the external feedback delay time $\tau_f$. Fig. 4 shows the temporal profiles of the number of plasmons (solid blue) and population inversion (dashed red) for the case of no drive (left column) and the corresponding phase space plots (right column) for the external feedback delay times $\tau_f = 100$ ((a), (e)), 200 ((b), (f)), 300 ((c), (g)), and 400 fs ((d), (h)) for $c_f = 0.5$. These plots show a high sensitivity to the feedback delay time which is reflected in varying the plasmon profiles and shapes of the phase plots.

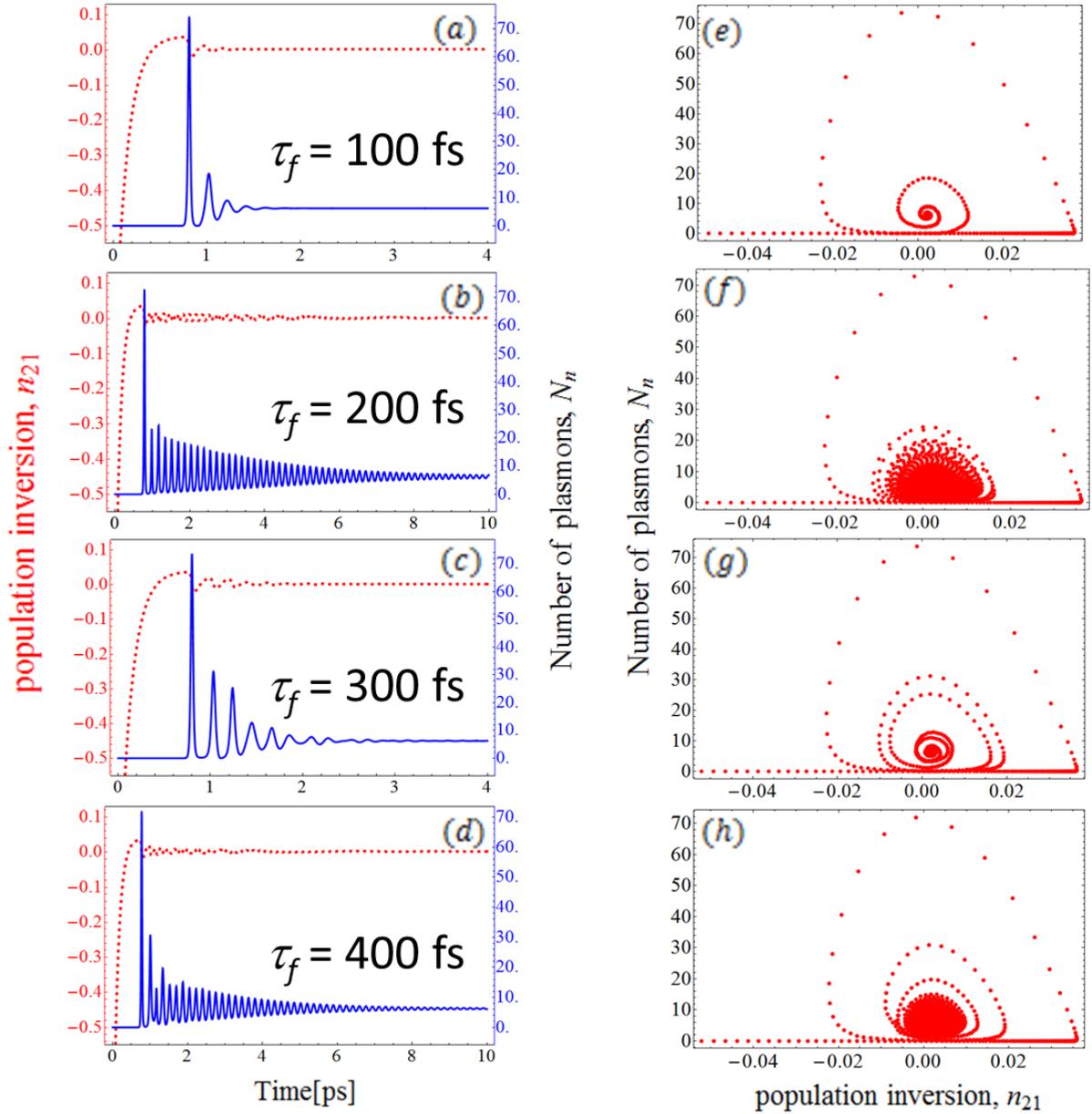

Figure 4. Effects of the feedback delay time without quantum coherence. Temporal profiles of the number of plasmons (solid blue) and population inversion (dashed red) for the case of no drive (left column) and the corresponding phase space plots (right column) for the external feedback delay times $\tau_f$ = 100 ((a), (e)), 200 ((b), (f)), 300 ((c), (g)), and 400 fs ((d), (h)) for $c_f = 0.5$.

Finally we perform similar simulations with a drive ($\Omega_a = 24 \times 10^{12}$ s$^{-1}$, Fig. 5) for $c_f = 0.5$. Fig. 5 shows the temporal profiles of the number of plasmons (solid blue) and population inversion (dashed red) for the case of the $\Omega_a = 24\times10^{12}$ s$^{-1}$ drive (left column) and the corresponding phase space plots (right column) for the external feedback delay times $\tau_f = 200$ ((a), (e)), 400 ((b), (f)), 600 ((c), (g)), and 800 fs ((d), (h)) for $c_f = 0.5$. These plots show a high sensitivity to the feedback delay time which is reflected in varying the plasmon profiles and shapes of the phase plots.

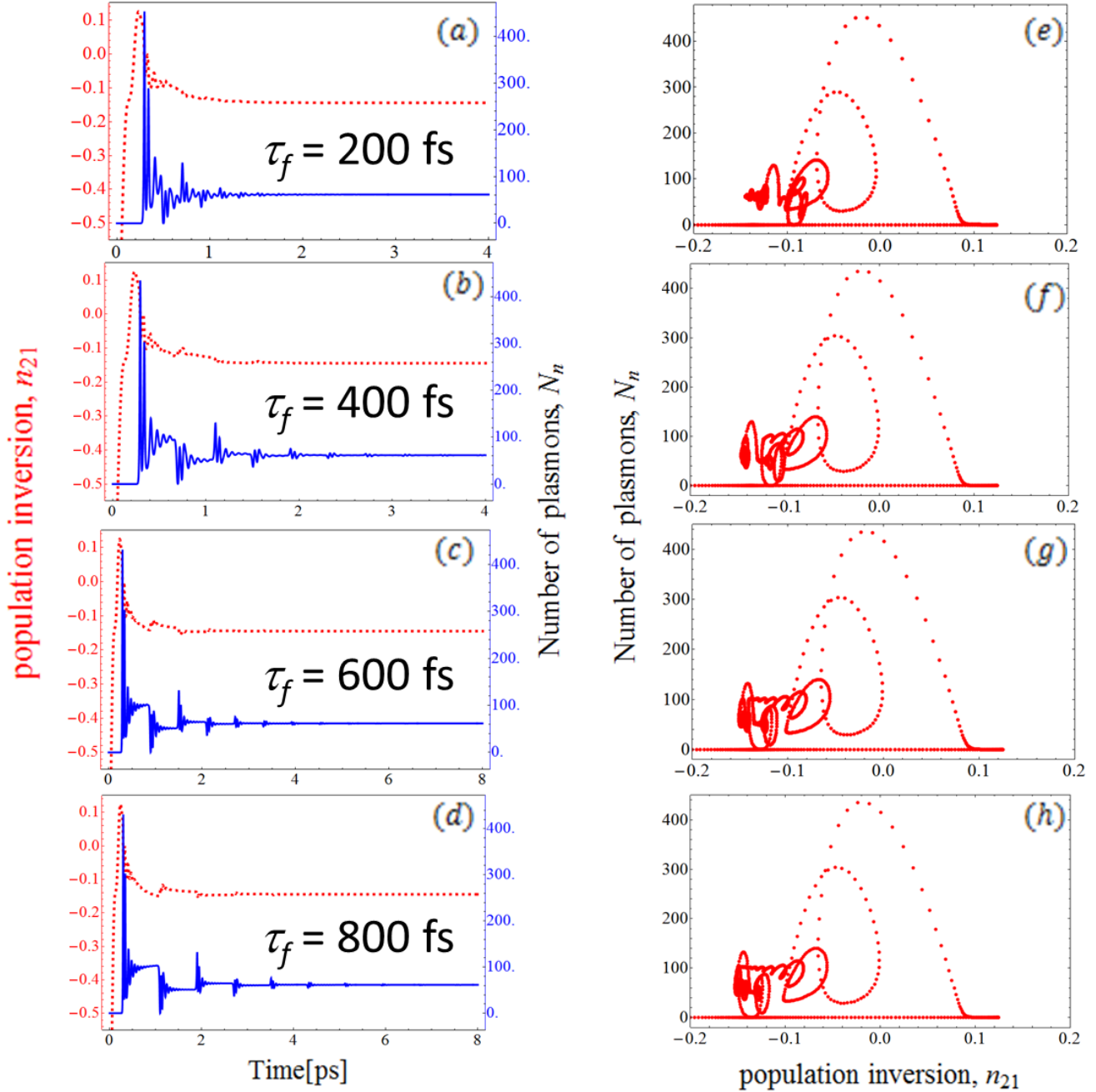

Figure 5. Effects of the feedback delay time with quantum coherence. Temporal profiles of the number of plasmons (solid blue) and population inversion (dashed red) for the case of the $\Omega_a = 24\times10^{12}$ s$^{-1}$ drive (left column) and the corresponding phase space plots (right column) for the external feedback delay times $\tau_f = 200$ ((a), (e)), 400 ((b), (f)), 600 ((c), (g)), and 800 fs ((d), (h)) for $c_f = 0.5$.

## 3. Discussion

Figs. 3 – 5 show the dependence of the temporal profiles of the spaser fields on the plasmonic feedback strength and delay time with and without the quantum coherence in the three-level gain medium induced by the laser drive. Fig. 3 shows that after a certain threshold coupling strength the effects of the feedback appear as periodic modulations of relaxation oscillation spikes. The threshold value of $c_f$ is lower in the presence of quantum coherence ($c_f \sim 0.1$) than in the absence ($c_f \sim 0.25$). Therefore, quantum coherence increases the sensitivity. For a high feedback strength without quantum coherence, the spaser can exhibit a chaotic behavior (Fig. 3d). The corresponding feedback with the quantum coherence does not show chaos. Quantum coherence significantly modifies the relaxation dynamics and provides a possibility to control chaos.

Our results can be used to extend the conventional feedback-based laser imaging techniques to the nanoscale. Figs. 4 and 5 show that the spaser dynamics is sensitive to the feedback delay time in both cases with and without coherence. However, the effects are different. Without coherence, the number of spikes varies with the feedback delay time but the periodic modulations are not visible for the delay times less than 1 ps. In the presence of quantum coherence, the modulations are visible for the delay time of 400 fs (Fig. 5b). This makes it possible to observe shorter plasmon propagation distances using the spaser relaxation dynamics. The shape of the phase plots in Figs. 4e – 4h vary with the delay time. However, they vary little in the presence of quantum coherence (Figs. 5e – 5h). The quantum-coherence-enhanced spaser dynamics is mainly influenced by the feedback delay time via the change of the relaxation spiking modulation frequency which is inversely proportional to the delay time.

Several experimental schemes are envisioned for controlling the feedback strength and delay time. For example, a spaser placed on a plasmonic metal surface will launch surface plasmon polaritons propagating on a surface and scattering off inhomogeneities or other plasmonic nanostructures. This is analogous to the laser radiation backscattered from an external target. The feedback strength may be controlled by varying the properties of the scattering nanostructures or of the surface. The feedback delay time may be controlled by varying the distance between the spaser and the backscattering nanostructure. The latter may be implemented using a scanning probe microscope tip which can be easily placed at varying distances from the spaser. The feedback strength may be controlled by varying the size and shape of the tip.

Spaser relaxation oscillations can be used to determine the one-way or round trip time of plasmon propagation. This can be used to develop other ultrafast nano-imaging techniques including surface plasmon polariton propagation[26], coherent surface-enhanced spectroscopy[27] and coherent multidimensional optical spectroscopy[28] and nanoscopy[29]. Resonant optical nanoantennas[30,31] may be optimized to improve the spaser performance.

**Acknowledgements**

We acknowledge the support of the National Science Foundation Grants No. EEC-0540832 (MIRTHE ERC), No. PHY-1068554, No. PHY-1241032 (INSPIRE CREATIV), and No. PHY-1307153, the Office of Naval Research, and the Robert A. Welch Foundation (Awards A-1261 and A-1547).

W.H. acknowledges the support of the Education Program for Talented Students of Xi'an Jiaotong University.